\documentclass[12pt]{article}
\usepackage{amsmath,amssymb}

\newcommand{\bflambda}{{\boldsymbol \lambda}}
\newcommand{\calB}{{\mathcal B}}
\newcommand{\calL}{{\mathcal L}}

\newcommand{\der}{\partial}

\begin{document}

\title{L\"owner equations and \\Dispersionless Hierarchies\footnote{%
Contribution to the Proceedings of XXIII International Conference of
Differential Geometric Methods in Theoretical Physics
Nankai Institute of Mathematics, Tianjin, China, August 2005.
}}

\author{Kanehisa Takasaki\footnote{\uppercase{W}ork partially
supported by \uppercase{G}rant-in-\uppercase{A}id for
\uppercase{S}cientific \uppercase{R}esearch by \uppercase{J}apan
\uppercase{S}ociety for the \uppercase{P}romotion of
\uppercase{S}cience,
\uppercase{N}o.\ 16340040.}
\\
Graduate School of Human and Environmental Sciences,\\ 
Kyoto University,\\
Kyoto 606-8502, Japan\\
E-mail: takasaki@math.h.kyoto-u.ac.jp
\\
\\
Takashi Takebe\footnote{\uppercase{W}ork partially
supported by \uppercase{G}rant-in-\uppercase{A}id for
\uppercase{S}cientific \uppercase{R}esearch by \uppercase{J}apan
\uppercase{S}ociety for the \uppercase{P}romotion of
\uppercase{S}cience,
\uppercase{N}o. 15540014.}
\\
Department of Mathematics, Ochanomizu University\\
Otsuka 2-1-1, Bunkyo-ku, Tokyo, 112-8610, Japan\\ 
E-mail: takebe@math.ocha.ac.jp}

\date{}

\maketitle

\abstract{
Reduction of a dispersionless type integrable system (dcmKP hierarchy) to
the radial L\"owner equation is presented.}

\section{Introduction}
\label{sec:intro}

Recently reductions and hodograph solutions of dispersionless/Whitham
type integrable systems are intensively
studied\cite{gib-tsa:99,yu-gib:00,m-a-m:02,g-m-a:03}. 
In this article we report another example; reduction of the
dispersionless coupled modified KP (dcmKP) hierarchy to the (radial)
L\"owner equation. 

The dcmKP hierarchy introduced by Teo\cite{teo:03-1} is an extension of
the dispersionless mKP hierarchy\cite{tak:02} with an additional degree
of freedom, or in other words, a ``half'' of the dispersionless Toda
lattice hierarchy\cite{tak-tak:91,tak-tak:95}.

The L\"owner equation was introduced by K.~L\"owner\cite{loe:23} in an
attempt to solve the Bieberbach conjecture. It is an evolution equation
of the conformal mapping from (a chain of) subdomains of the unit disk
onto the unit disk. We can also define the same kind of equation with
different normalization which is called the ``{\em chordal} L\"owner
equation''. See Lawler, Schramm and Werner\cite{l-s-w:01} \S2.3 for
details. The original L\"owner equation is, therefore, often called the
``{\em radial} L\"owner equation''.

The reduction of the dispersionless KP
hierarchy\cite{dkp,tak-tak:92,tak-tak:95} to the chordal L\"owner
equation (and its generalization) has been studied by Gibbons and
Tsarev\cite{gib-tsa:99}, Yu and Gibbons\cite{yu-gib:00}, Ma\~nas,
Mart\'{\i}nez Alonso and Medina\cite{m-a-m:02} and others. Our question
is: how about the radial L\"owner equation? The answer is that there
appears another degree of freedom and the resulting system turns out to
be the dcmKP hierarchy.

In the following two sections we review the two ingredients, the
L\"owner equation and the dcmKP hierarchy. The main result is presented
in the last section. Details including proofs will be published in the
forthcoming paper.

\section{Radial L\"owner equation}
\label{sec:loewner}

In this section we review the (radial) L\"owner equation and introduce
related notions. Since we are interested in algebro-analytic nature of
the system, we omit reality/positivity conditions which are essential in
the context of the complex analysis.

The L\"owner equation is a system of differential equations for a
function 
\begin{equation}
    w= g(\bflambda,z) 
    = e^{-\phi(\bflambda)} z 
    + b_0(\bflambda) 
    + b_1(\bflambda) z^{-1} + b_2(\bflambda) z^{-2} + \cdots 
\label{g(lam;z)}
\end{equation}
where $\bflambda=(\lambda_1,\dots,\lambda_N)$ and $z$ are independent
variables. In the complex analysis the variable $z$ moves in a subdomain
of the compliment of the unit disk and the variables $\lambda_i$
parametrize the subdomain. In our context $g(\bflambda,z)$ is considered
as a generating function of the unknown functions $\phi(\bflambda)$ and
$b_n(\bflambda)$. We assume that for each $i=1,\dots,N$ a {\em driving
function} $\kappa_i(\bflambda)$ is given. The {\em L\"owner equation} is
the following system:
\begin{equation}
    \frac{\der g}{\der \lambda_i}(\bflambda;z)
    = g(\bflambda;z) 
    \frac{\kappa_i(\bflambda) + g(\bflambda;z)}
         {\kappa_i(\bflambda) - g(\bflambda;z)}
      \frac{\der \phi(\bflambda)}{\der \lambda_i}, \qquad
    i=1,\dots,N.
\label{r-loewner:g}
\end{equation}
(The original L\"owner equation\cite{loe:23} is the case $N=1$.)

Later the inverse function of $g(\bflambda,z)$ with respect to the
$z$-variable will be more important than $g$ itself. We denote it by
$f(\bflambda,w)$:
\begin{equation}
    z = f(\bflambda,w)
    = e^{\phi(\bflambda)} w 
    + c_0(\bflambda)
    + c_1(\bflambda) w^{-1} + c_2(\bflambda) w^{-2} + \cdots.
\label{f(lambda;w)}
\end{equation}
It satisfies $g(\bflambda,f(\bflambda,w)) = w$ and
$f(\bflambda,g(\bflambda,z)) = z$, from which we can determine the
coefficients $c_n(\bflambda)$ in terms of $\phi(\bflambda)$ and
$b_n(\bflambda)$. The L\"owner equation \eqref{r-loewner:g} is rewritten
as the equation for $f(\bflambda,w)$ as follows:
\begin{equation}
    \frac{\der f}{\der \lambda_i}(\bflambda;w)
    = w
    \frac{w + \kappa_i(\bflambda)}{w - \kappa_i(\bflambda)}
    \frac{\der \phi(\bflambda)}{\der \lambda_i}
    \frac{\der f}{\der w}(\bflambda;w).
\label{r-loewner:f}
\end{equation}
This equation leads to the compatibility condition of $\kappa_i$'s:
\begin{align}
    \frac{\der \kappa_j}{\der \lambda_i}
    &= - \kappa_j
    \frac{\kappa_j + \kappa_i}{\kappa_j-\kappa_i} 
    \frac{\der \phi}{\der \lambda_i},
\label{dk/dlambda}
\\
    \frac{\der^2 \phi}{\der\lambda_i \der\lambda_j}
    &=
    \frac{4 \kappa_i \kappa_j}{(\kappa_i - \kappa_j)^2}
    \frac{\der \phi}{\der \lambda_i}\, 
    \frac{\der \phi}{\der \lambda_j},
\label{d2phi/dlambda2}
\end{align}
for any $i,j$ ($i\neq j$).

The {\em Faber polynomials} are defined as follows\cite{teo:03-2}:
\begin{equation}
    \Phi_n(\bflambda,w) := (f(\bflambda,w)^n)_{\geq 0}.
\label{def:faber}
\end{equation}
Here $(\cdot)_{\geq 0}$ is the truncation of the Laurent series in $w$
to its polynomial part.

\section{dcmKP hierarchy}
\label{sec:dcmkp}

We give a formulation of the dcmKP hierarchy different from
Teo\cite{teo:03-1}. The equivalence (up to a gauge factor) will be
explained in the forthcoming paper.

The independent variables of the system is $(s,x,t)$ where
$t=(t_1,t_2,\dots)$ is a series of infinitely many variables. The
variables $x$ and $t_1$ appear in the equations only as the combination
$x+t_1$, so we often omit $x$. Namely, ``$t_1$'' should be understood as
the abbreviation of $x+t_1$. The unknown functions $\phi(s,t)$ and
$u_n(s,t)$ ($n=0,1,2,\dots$) are encapsulated in the series
\begin{equation}
    \calL(s,t;w) 
    = e^{\phi(s,t)} w 
    + u_0(s,t) + u_1(s,t) w^{-1} + u_2(s,t) w^{-2} + \cdots,
\label{def:L:dcmkp}
\end{equation}
where $w$ is a formal variable. The {\em dispersionless coupled modified
KP hierarchy} (dcmKP hierarchy) is the following system of differential
equations: 
\begin{equation}
    \frac{\der\calL}{\der t_n} = \{\calB_n, \calL\}, \qquad
    n=1,2,\dots.
\label{dcmkp}
\end{equation}
Here the Poisson bracket $\{,\}$ is defined by
\begin{equation}
    \{f(s,x),g(s,x)\} :=
    w \frac{\der f}{\der w} \frac{\der g}{\der s} -
    w \frac{\der f}{\der s} \frac{\der g}{\der w},
\label{def:poisson}
\end{equation}
and $\calB_n$ is the polynomial in $w$ defined by
\begin{equation}
    \calB_n
    := (\calL^n)_{>0} + \frac{1}{2} (\calL^n)_0,
\label{def:Bn:dcmkp}
\end{equation}
where $(\cdot)_{>0}$ is the positive power part in $w$ and $(\cdot)_0$
is the constant term with respect to $w$.

It is easy to construct a theory for this system similar to those for
the dispersionless KP hierarchy or the dispersionless Toda
hierarchy\cite{tak-tak:95}. 

\section{Main results}
\label{sec:results}

In this section we show that a specialization of the variables
$\bflambda$ in $f(\bflambda,w)$ gives a solution of the dcmKP
hierarchy. 

Suppose $\bflambda(s,t) = (\lambda_1(s,t), \dots, \lambda_N(s,t))$
satisfies the equations
\begin{equation}
    \frac{\der \lambda_i}{\der t_n}
    = v^n_i(\bflambda(s,t))
    \frac{\der \lambda_i}{\der s},
\label{lambda(s,t)}
\end{equation}
where $v^n_j(\bflambda)$ are defined by
\begin{equation}
    v^n_j(\bflambda)
    := \kappa_j(\bflambda)
    \frac{\der \Phi_n}{\der w}(\bflambda,\kappa_j(\bflambda))
    =
    \left.\frac{\der \Phi_n}{\der \log w} (\bflambda, w)
    \right|_{w=\kappa_j(\bflambda)}.
\label{def:vnj}
\end{equation}
They satisfy the equations
\begin{equation}
    \frac{\der v^n_j}{\der \lambda_i}
    =
    V_{ij} (v^n_i-v^n_j),
\label{dv/dlambda}
\end{equation}
where 
\begin{equation}
    V_{ij} :=
    \frac{2 \kappa_i \kappa_j}{(\kappa_i-\kappa_j)^2}
    \frac{\der \phi}{\der \lambda_i}.
\label{def:Vij}
\end{equation}
The hydrodynamic type equations \eqref{lambda(s,t)} can be solved by the
generalized hodograph method of Tsarev\cite{tsa:91}: Let
$F_i(\bflambda)$ be functions satisfying 
\begin{equation}
    \frac{\der F_j}{\der \lambda_i}
    =
    V_{ij} (F_i - F_j).
\label{Fi:cond}
\end{equation}
Then the {\em hodograph relation}
\begin{equation}
    F_i(\bflambda(s,t)) 
    = s + \sum_{n=1}^\infty v^n_i(\bflambda(s,t)) \, t_n
\label{hodograph}
\end{equation}
determines the solution of \eqref{lambda(s,t)}, $\bflambda(s,t)$, as the
implicit function.

Our main result is as follows: let $f(\bflambda,w)$ be a solution of the
radial L\"owner equation \eqref{r-loewner:f} of the form
\eqref{f(lambda;w)} and $\bflambda(s,t)$ be a solution of
\eqref{lambda(s,t)}. Then the function $\calL=\calL(s,t;w)$ defined by
\begin{multline}
    \calL(s,t;w) := f(\bflambda(s,t), w)
\\    = e^{\phi(\bflambda(s,t))} w 
    + c_0(\bflambda(s,t))
    + c_1(\bflambda(s,t)) w^{-1} + c_2(\bflambda(s,t)) w^{-2} + \cdots
\label{def:L}
\end{multline}
is a solution of the dcmKP hierarchy \eqref{dcmkp}.

In the proof we construct the $S$-function\cite{tak-tak:91,tak-tak:95},
following the method by Ma\~nas-Alonso-Medina\cite{m-a-m:02}. 

If we start from the chordal L\"owner equation instead of the radial
L\"owner equation, we obtain a solution of the dispersionless KP
hierarchy. This is due to Gibbons and Tsarev\cite{gib-tsa:99}, Yu and
Gibbons\cite{yu-gib:00}, Ma\~nas, Mart\'{\i}nez Alonso and
Medina\cite{m-a-m:02}. The generalization to the Whitham hierarchies is
considered by Guil, Ma\~nas and Mart\'{\i}nez
Alonso\cite{g-m-a:03}. Note that their generalization does not contain
the radial L\"owner case, because of the normalization at the infinity.

\end{document}